\documentclass[journal]{IEEEtran}
\ifCLASSINFOpdf
  \usepackage[pdftex]{graphicx}
\else
\fi
\usepackage{cite}
\usepackage{amsmath}
\usepackage{amsfonts} 

\usepackage{algorithmic}
\usepackage{array}
\usepackage{color}
\usepackage{subfigure}
\usepackage{mdwtab}
\usepackage{booktabs}
\usepackage{multirow}
\usepackage{hyperref}
\usepackage{threeparttable}

\begin{document}
\title{Kernel Learning for High-Resolution Time-Frequency Distribution}

\author{Lei~Jiang, Haijian~Zhang, Lei~Yu, and Guang~Hua 

\thanks{All the authors are with Signal Processing Lab., School of Electronic Information, Wuhan University, China. } 
\vspace{-1.\baselineskip}
}

\markboth{TIME-FREQUENCY ANALYSIS AND ITS APPLICATIONS}{}
\maketitle

\begin{abstract}

The design of high-resolution and cross-term (CT) free time-frequency distributions (TFDs) has been an open problem. Classical kernel based methods are limited by the trade-off between TFD resolution and CT suppression, even under optimally derived parameters. To break the current limitation, we propose a data-driven kernel learning model directly based on Wigner-Ville distribution (WVD). The proposed kernel learning based TFD (KL-TFD) model includes several stacked multi-channel learning convolutional kernels. Specifically, a skipping operator is utilized to maintain correct information transmission, and a weighted block is employed to exploit spatial and channel dependencies. These two designs simultaneously achieve high TFD resolution and CT elimination.  Numerical experiments on both synthetic and real-world data confirm the superiority  of the proposed KL-TFD over traditional kernel function methods.

\end{abstract}

\begin{IEEEkeywords}
Kernel learning, kernel function, high-resolution time-frequency distribution.
\end{IEEEkeywords}
\IEEEpeerreviewmaketitle

\section{Introduction}

\IEEEPARstart{T}{ime-frequency} distributions (TFDs) provide a comprehensive description of nonstationary signals \cite{2013LjubisaStankovic,2016ivBoualem,2018CambridgePatrick,saulig2019extraction,Zhang9472949},  but their performances are restrained by issues such as time-frequency (TF)  resolution, cross-terms (CTs), noise, spectrally overlapped components,  etc. Therefore, high-resolution and CT-free TFDs \cite{7401107Zuo,7736146Zuo} to tackle the above issues are of vital importance for practical applications as most real-world signals are nonstationary and random \cite{Vaishali9443566,Zhang9237164,flandrin2010time,7073495Zhang,ZHANG2015141,SEJDIC2009153,volaric2017data}.

Generally, linear TFDs suffer from low TF resolution, although some new  linear TFDs have been explored \cite{yu2017synchroextracting,Yu8458385,abdoush2019adaptive}. Quadratic TFDs, e.g., Wigner-Ville distribution (WVD) \cite{2016ivBoualem,Mohammad9456035}, have high TF resolution,  but it is difficult  for them to interpret the signal in 2-dimensional (2D) TF plane due to significant CTs \cite{127284Hlawatsch}. Considerable investigations are therefore dedicated to reducing CTs while preserving high TF concentration. Particularly, kernel function based TFD methods viewed as smoothed versions of WVD have been studied \cite{BOASHASH2018120}, and they can be divided into two categories according to whether they depend on signals. Since CTs have oscillatory characteristics, signal-independent kernel functions are essentially low-pass filters, e.g., Choi–Williams distribution (CWD) \cite{28057} and B-distribution (BD) \cite{950779}. Moreover, the kernels separately smoothing along both time and frequency axes are designed to further improve TF resolution, e.g., S-method (SM) \cite{258146}, extended modified B-distribution (EMBD) \cite{boashash2013time}, and compact kernel distribution (CKD) \cite{6165379Abed}.

However, the above signal-independent kernels require fixed or manually selected parameters, which might ignore some crucial information about signal's intrinsic characteristics, e.g., the direction of analytic signals. Thus, signal-dependent kernels are designed with respect to certain criteria to gain better TF representation, e.g., radially Gaussian kernel (RGK) \cite{baraniuk1993signal}, adaptive optimal-kernel (AOK) \cite{319606,4698541995},  multi-directional distribution (MDD) \cite{boashash2017improved}, and adaptive directional TFD (ADTFD) \cite{khan2016multi,s000349,mohammadi2018locally}. Although most of signal-dependent kernels achieve better performance than signal-independent ones, they need to know the signal type and solve a kernel optimization problem subject to certain criteria. In addition, both signal independent and dependent kernel based methods have the limitation to simultaneously improve the performance of  TF resolution and CT reduction. The above difficulties inspire us to develop new methods, aiming to break the trade-off between high TF resolution and negligible CTs.

Motivated by the 2D convolution relationship between WVD and smoothed TFDs, this letter proposes a kernel learning based TFD (KL-TFD) directly from WVD.  We replace traditional kernel functions by designing a convolutional neural network (CNN), where multi-channel learning convolutional kernels are utilized to simulate traditional kernel functions.

\section{The Proposed Kernel Learning Based TFD}
\begin{figure*}[!t]
\centering
\includegraphics[width=6.2 in]{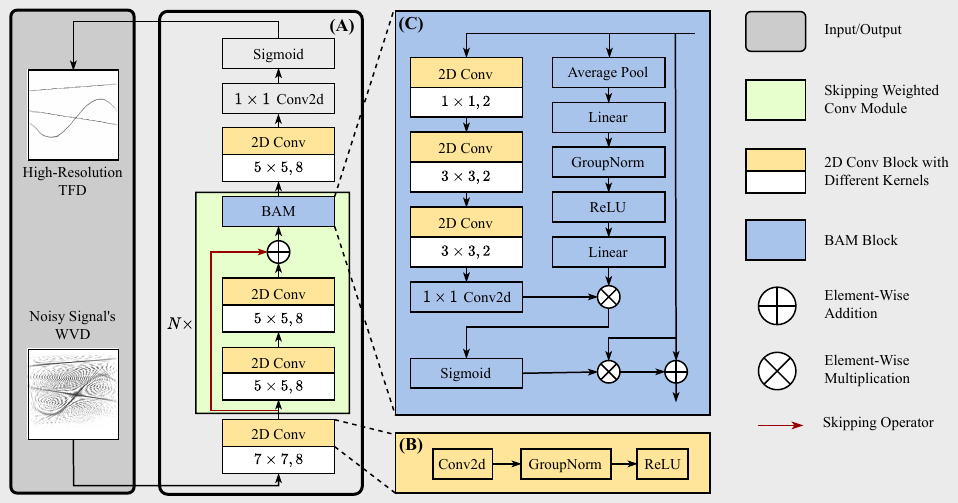}
\vspace{-0mm}
\caption{Schematic diagram of the proposed kernel learning based TFD model. (A) The KL-TFD  with $N$ skipping weighted Conv modules, where the skipping operator is indicated by the red line. (B) The structure of the 2D Conv block. (C) The structure of the weighted block via BAM.} \label{fig1}
\end{figure*}

The architecture of the proposed KL-TFD model is shown in Fig. \ref{fig1} (A), where a 2D Conv block is firstly used on signal's WVD to obtain a multi-channel input. The KL-TFD mainly includes $N$ stacked skipping weighted Conv modules, each of which consists of two 2D Conv blocks and one weighted block.  The 2D Conv blocks with varying kernel sizes have the same structure, as shown in Fig. \ref{fig1} (B). The weighted block  is realized by the bottleneck attention module (BAM) \cite{DBLP:journals/corr/abs-1807-06514,Park2020ijcv} to learn spatial and channel dependencies, as illustrated in Fig. \ref{fig1} (C). Finally,  a 2D Conv block and a $1\times1$ convolutional kernel convert multi-channel into single-channel output.

\vspace{-1.5mm}
\subsection{The KL-TFD Model}
The ambiguity function (AF) of an analytic signal $z(t)$ is defined as:
\begin{align} A_{z}(\theta, \tau)= \int_{-\infty}^{\infty} z\left(t+\frac{\tau}{2}\right) z^{*}\left(t-\frac{\tau}{2}\right) e^{-j 2 \pi\theta t} \mathrm{d} t,
\label{eq1}
\end{align}
where $\theta$ is the frequency shift, and $\tau$ is the time lag. Since Cohen's class distributions can be interpreted as weighted versions of AF smoothed by a low-pass filter $\Phi(\theta, \tau)$, the general formula of Cohen's class TFDs is expressed as \cite{BOASHASH2018120}: 
\begin{align} 
\zeta_{z}(t, f)=\int_{-\infty}^{\infty} \int_{-\infty}^{\infty} A_{z}(\theta, \tau) \Phi(\theta, \tau)  e^{j 2 \pi(\theta t-f \tau)} \mathrm{d} \theta \mathrm{d} \tau.
\label{eq2}
\end{align}
When $\Phi(\theta, \tau)=1$, $\zeta_{z}(t, f)$ is degenerated into WVD, which is the 2D Fourier transform of AF, i.e.,
\begin{align}
W_{z}(t, f)=\int_{-\infty}^{\infty} z\left(t+\frac{\tau}{2}\right) z^{*}\left(t-\frac{\tau}{2}\right) e^{-j 2 \pi f \tau} \mathrm{d} \tau .
\label{eq1b}
\end{align}

According to the convolution theorem and (\ref{eq1b}), the expression in equation (\ref{eq2}) can be rewritten as the form of 2D convolution operation as below \cite{BOASHASH2018120}:
\begin{align} \label{eq4}
\zeta_{z}(t, f) = W_{z}(t, f) * * \gamma(t, f),
\end{align}
where $* *$ denotes the 2D convolution operator, and $\gamma(t, f)$ is a 2D convolutional kernel, which is computed by 2D Fourier transform of the AF kernel function $\Phi(\theta, \tau)$. Additionally, the ADTFD is defined as the 2D convolution of a quadratic TFD with an adaptive directional filter \cite{s000349}:
\begin{align}\label{eq5}
\zeta_{\text{adapt}}(t, f)=W_{z}(t, f) * * \gamma_{\beta}(t, f),
\end{align}
where $\gamma_{\beta}(t, f)$ denotes the double derivative directional Gaussian filter, and $\beta$ is the rotation angle with respect to the time axis. For each TF point, its angle is selected by solving an optimization problem over a group of directions \cite{s000349}.

Inspired by the 2D convolution relationship in (\ref{eq4}) and (\ref{eq5}), we propose to employ multi-channel learning convolutional kernels to replace kernel functions or adaptive directional filters. As seen in Fig. \ref{fig1} (A), the signal's WVD  serves as the input data,  followed by $N$ stacked skipping weighted Conv modules to obtain resolution improvement and CT elimination.

\vspace{-1.5mm}
\subsection{The 2D Conv Block Design with Skipping Operator}
Each skipping weighted Conv module contains two 2D Conv blocks and a weighted block. Fig. \ref{fig1} (B) shows the detail of a 2D Conv block, which involves a convolutional kernel followed by group normalization together with the ReLU function. It is examined that a larger kernel size leads to a more smooth TFD, we thus set the kernel sizes in the two 2D Conv blocks with $5 \times 5$, and  the number of filters  are both set as $C=8$. Since useful information may as well be transmitted with degradation in network, a skipping operator is used to maintain correct information transmission.  
Assuming the number of frequency samples is equal to that of time samples. For a  $K$-sample signal,  given the output of the $n$-th skipping weighted Conv module $\zeta^{(n)}_{\text{sw}} \in \mathbb{R}^{C \times K \times K}$, we have the output of 2D Conv blocks after the  $n$-th skipping operator:
\begin{align} \label{spleq6}
\zeta^{(n)}_{\text{s}}(t, f)=\zeta^{(n-1)}_{\text{sw}}(t, f)* *\gamma_{s_{n_1}}* *\gamma_{s_{n_2}} + \zeta^{(n-1)}_{\text{sw}}(t, f),
\end{align}
where $n=1,2, \cdots, N$ and $\zeta^{(0)}_{\text{sw}}(t, f)$ is the multi-channel WVD input (output of the $7\times 7$ 2D Conv block). $\gamma_{s_{n_1}}$ and $\gamma_{s_{n_2}}$ denote the kernels corresponding to the two 2D Conv blocks  in the $n$-th skipping weighted Conv module.

\vspace{-1mm}
\subsection{The Weighted Block Design via BAM}
To properly determine the direction of each TF point, element-wise weights are incorporated in a way such that primary directions are assigned with large weights, while insignificant directions are assigned with small ones. The directional filer in ADTFD adaptively selects directions by optimizing the correlation between the directional kernel and the modulus of TFD. In contrast, the data-driven kernel learning model can benefit from sufficient training data and our end-to-end network structure tends to be easy to train. To be specific, we adopt the same network architecture as the BAM \cite{DBLP:journals/corr/abs-1807-06514} to  utilize dependencies among TF coefficients as well as different channels, which is illustrated in Fig. \ref{fig1} (C). Spatial dependencies can further distinguish auto-terms (ATs) from CTs as ATs and CTs have different directional features. Channel attention jointly attends to information from different representation subspaces. 
In order to reduce the amount of parameters, the 2D Conv blocks in BAM have smaller kernel sizes compared with those of skipping 2D Conv blocks. Channel and spatial weights are two separate branches, thus element-wise weights are computed as:
\begin{align} \label{spleq7}
\boldsymbol{\omega}_{n}\!=\!\sigma\Big(\mathbf{F}_{ch}\big (\zeta^{(n)}_{\text{s}}\big) \otimes \mathbf{F}_{sp}\big(\zeta^{(n)}_{\text{s}}\big)\Big),
\end{align}
where $n=1,2, \cdots, N$ and $\boldsymbol{\omega}_{n} \in \mathbb{R}^{C \times K \times K}$, $\otimes$ denotes the element-wise multiplication. Channel weights are defined as:
\begin{align} \label{spleq8}
\mathbf{F}_{ch}(\boldsymbol{\rho}_{n})=\sigma\big(g(\boldsymbol{\rho}_{n}, \mathbf{W})\big)\!=\!\sigma\big(\mathbf{W}_{2} \delta\left(\mathbf{W}_{1}\boldsymbol{\rho}_{n} \right)\big),
\end{align}
where $\mathbf{F}_{ch}(\boldsymbol{\rho}_{n}) \in \mathbb{R}^{C \times K \times K}$, and $\boldsymbol{\rho}_{n} \in \mathbb{R}^{C \times 1 \times 1} $ is the output of average pooling operator over $\zeta^{(n)}_{\text{s}}$. $g(x, \mathbf{W})$ represents a gating mechanism, which forms a bottleneck with two fully-connected layers around the non-linearity. The first fully-connected layer $\mathbf{W}_{1} \in \mathbb{R}^{C \times{C}/{r_{c}}}$ achieves channel reduction with reduction ratio $r_c$, then non-linearity characteristic is introduced by $\delta$ (ReLU function), and the second fully-connected layer $\mathbf{W}_{2} \in \mathbb{R}^{{C}/{r_{c}} \times C}$ increases the number of channels. Finally, channel weights are obtained by Sigmoid function $\sigma (x)= {1}/\big({1+\exp(-x)}\big)$. On the other hand, spatial weights are attained by a bottleneck module with four convolutional layers using a reduction ratio $r_s$:
\begin{align} \label{spleq9}
\mathbf{F}_{sp}\big(\zeta^{(n)}_{\text{s}}\big)\!=\!\zeta^{(n)}_{\text{s}}(t, f)\!* *\gamma_{w_{n_1}}\!* *\gamma_{w_{n_2}}\!* *\gamma_{w_{n_3}}* *\gamma_{w_{n_4}},
\end{align}
where $\mathbf{F}_{sp}\big(\zeta^{(n)}_{\text{s}}\big) \in \mathbb{R}^{C \times K \times K}$, and $\{\gamma_{w_{n_i}}\}_{i=1,2,3,4}$ denote the kernels of the four Conv blocks in the $n$-th weighted block.

As a result, the weighted output in the $n$-th skipping weighted Conv module can be obtained via (\ref{spleq6}) to (\ref{spleq9}):
\begin{align} \label{spleq10}
\zeta^{(n)}_{\text{sw}}(t, f)=\boldsymbol{\omega}_{n} \otimes \zeta^{(n)}_{\text{s}}(t, f) + \zeta^{(n)}_{\text{s}}(t, f),
\end{align}
where $n=1,2,\cdots,N$, and element-wise weights realize proper direction selection at each TF point. On the basis of the 2D Conv block with skipping operator, the BAM can attain channel and spatial dependencies as well, which further improves TF resolution and eliminates CTs.

From the viewpoint of parameter reduction, we make the choice of a 2D Conv block and a $1 \times 1$ convolutional kernel to conduct channel fusion on $\zeta^{(N)}_{\text{sw}} \in \mathbb{R}^{C \times K \times K}$ in (\ref{spleq10}), and the sum of all channels  leads to a  high-resolution and CT-free TFD, as illustrated in Fig. \ref{fig1}. 

\begin{figure*}[!h]
\begin{center}
\hspace{-3mm}
\includegraphics[width=6.5 in]{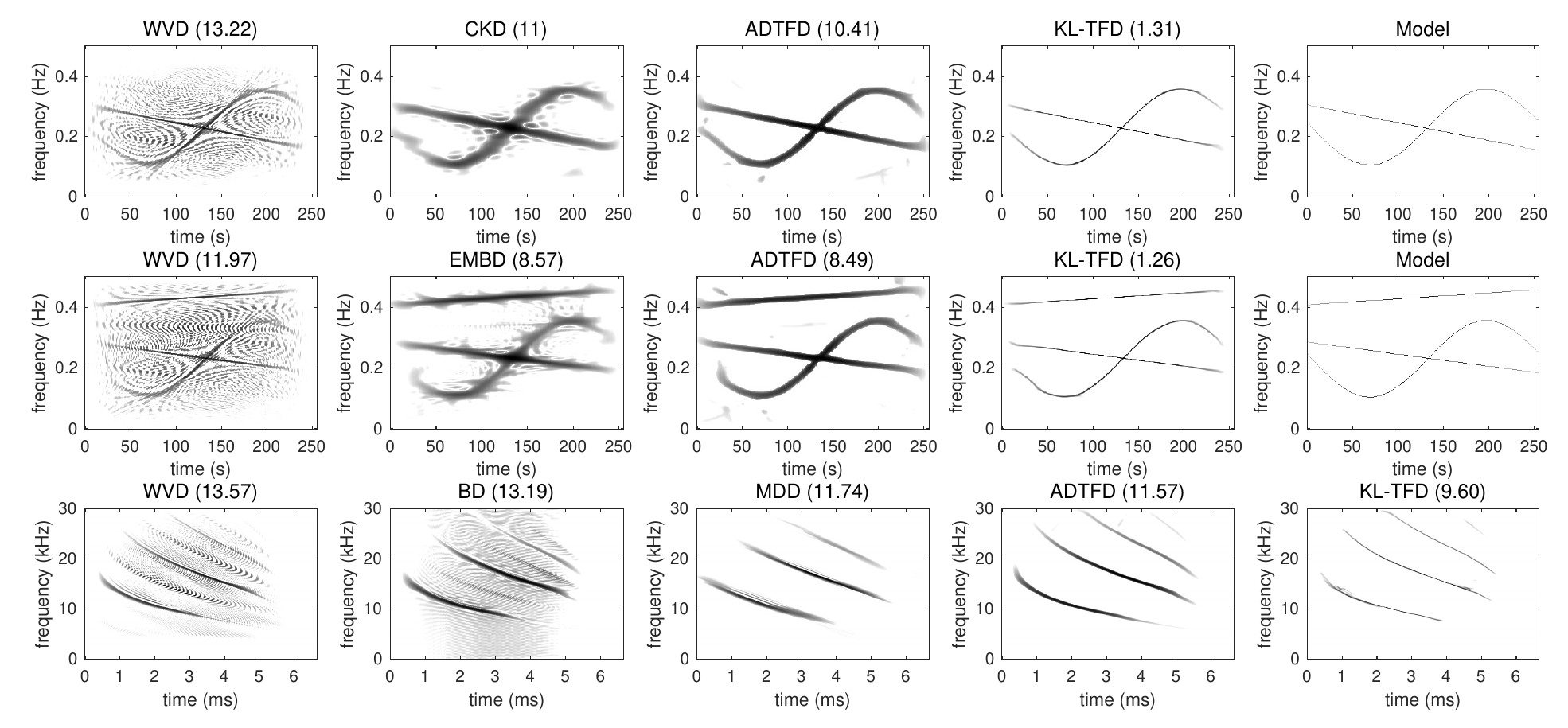}
\vspace{-2mm}
\caption{Generated TFDs of synthetic and real-world data using WVD \cite{2016ivBoualem},  BD \cite{950779}, EMBD\cite{boashash2013time}, MDD \cite{boashash2017improved}, CKD \cite{6165379Abed}, ADTFD \cite{khan2016multi}, and  the proposed  KL-TFD. \textbf{(First row)}: TFDs of a synthetic signal including an AM-LFM and an AM-SFM at SNR = 10 dB. The \textit{ $\ell_1$ distance to model} is annotated on the top of each TFD. \textbf{(Second row)}: TFDs of a synthetic signal including two AM-LFM and an AM-SFM at SNR = 10 dB. The \textit{ $\ell_1$ distance to model} is annotated on the top of each TFD. \textbf{(Last row)}: TFDs of a real-world bat echolocation signal. The \textit{ R\'{e}nyi entropy} is annotated on the top of each TFD.}\label{fig2}
\label{fig2}
\end{center}
\end{figure*}

\vspace{-1mm}
\section{Numerical Experiments}

The proposed KL-TFD model is compared with state-of-the-art kernel design methods \cite{BOASHASH2018120,mohammadi2018locally}, including WVD \cite{2016ivBoualem}, CKD \cite{6165379Abed}, BD \cite{950779}, EMBD \cite{boashash2013time}, MDD \cite{boashash2017improved}, AOK \cite{4698541995}, RGK \cite{baraniuk1993signal}, and ADTFD \cite{khan2016multi}. The \textit{$\ell_1$ distance to model} and \textit{R\'{e}nyi entropy} \cite{STANKOVIC2001621,flandrin2010time} are regarded as two criteria to quantitatively assess the above-mentioned  methods. For the parameter setting of the KL-TFD, channel reduction and  spatial reduction in weighted blocks are set to $r_c=r_s=4$. Our training dataset including $8\times 10^4$ TF images is built by synthetic signal mixtures, each of which is constituted of two or three randomly spectrally-overlapped (only one intersection) linear frequency-modulated (LFM) and sinusoidal frequency-modulated (SFM) components with amplitude modulation (AM) at a fixed SNR = 10 dB. We train the KL-TFD network on the above training dataset using NVIDIA GeForce GTX 1080 GPUs with step decay learning rate schedule. Commonly-used binary cross entropy (BCE) loss function is adopted to obtain a robust network.  Table \ref{tab3} presents the \textit{$\ell_1$ distance to model} results on a three-component synthetic signal (shown in the second row of Fig. \ref{fig2}) at different SNR levels, implemented by our KL-TFD with varying number of the skipping weighted Conv modules, i.e., $N\in\{3,5,7,9,11,13,15\}$. Note that the performance improves as the value of $N$ increases when SNR $\geq$ 10 dB, and becomes slightly inconsistent when SNR $<$ 10 dB. The reason behind
this issue is that we only train our model with data at SNR
= 10 dB.  To balance the performance and complexity in moderate-to-high SNR cases, we choose SNR = 10 dB for training and set $N=9$ in the following experiments. One can build specific training data depending on practical requirements. For more information, please refer to  \href{https://github.com/teki97/KL-TFD}{https://github.com/teki97/KL-TFD}.

 \begin{table}[!t]
\center
\begin{threeparttable}[b]
\renewcommand{\arraystretch}{1.}
 \caption{$\ell_1$ distance to model using our KL-TFD with increasing number of the skipping weighted Conv modules.}\label{tab3}
 \setlength{\tabcolsep}{1.1mm}{
 \begin{tabular}{ l ccccccc}
  \toprule\toprule
SNR     & $N=3$ & $N=5$  & $N=7$   & $N=9$  & $N=11$  &  $N=13$ & $N=15$ \\
  \midrule
45 dB & 1.60 & 1.53 & 1.22 & 1.19     &1.15 & 1.15 & 1.13\\
35 dB & 1.60& 1.53 & 1.22 & 1.19   &1.15 & 1.15 & 1.13\\
25 dB & 1.61 & 1.53 & 1.22 & 1.19   &1.15 & 1.15  & 1.14\\
15 dB & 1.63 & 1.57 & 1.26 & 1.23   &1.20 & 1.20  & 1.18\\
10 dB & 1.69 & 1.63 & 1.33 & 1.32 & 1.29 & 1.28 & 1.28\\
5 dB  & 1.85 & 1.74 & 1.47 & 1.52 & 1.29 & 1.48 & 1.48\\
 0 dB & 2.27 & 1.87 & 1.72 & 1.98  &1.46 &1.81 & 1.82\\
  \bottomrule
 \end{tabular}
 }
  \end{threeparttable}\vspace{-2mm}
  \end{table}

\vspace{-1mm}
\subsection{TFDs of Synthetic Spectrally-Overlapped Signals}

We consider a 256-sample two-component test signal containing an AM-LFM and an AM-SFM, as well as a 256-sample three-component test signal with two AM-LFM and an AM-SFM. The generated TFDs on these two synthetic signals using different TFD methods with SNR = 10 dB  are shown in the first and second rows of  Fig. \ref{fig2}. It is seen that the WVD is seriously disturbed by both CTs and noises. The CKD and EMBD suppress the noises at the sacrifice of TF resolution. Although the ADTFD can greatly eliminate CTs without TFD distortion by considering directions of  TF points, it is crucial to make a balance between TF resolution and CT reduction.  Besides, the parameters of  filter shape and window length for the ADTFD are manually selected. Experimental results show that the proposed KL-TFD method can break the trade-off between high TF resolution and CT suppression, which is an issue often encountered by traditional TFDs.

\begin{table}[!h]
\center
\begin{threeparttable}[b]
\renewcommand{\arraystretch}{1.}
\vspace{-2mm}
 \caption{$\ell_1$ distance to model using different methods on the  three-component AM-FM signal at various SNR levels.\label{tab1}}
 \setlength{\tabcolsep}{2.3mm}{
 \begin{tabular}{ l cccccc}
  \toprule\toprule 
  SNR      & WVD  & EMBD   & AOK  & ADTFD & RGK  & \textbf{KL-TFD} \\
  \midrule
45 dB & 9.76 & 8.07 & 7.44  & 8.26 &4.67   & \textbf{1.19}\\
35 dB & 9.77 & 8.08 & 7.43 & 8.27 &4.67  &\textbf{1.19}\\
25 dB & 9.90 & 8.09 & 7.45 & 8.27 &4.69    & \textbf{1.19}\\
15 dB & 10.75 & 8.40 & 7.61 & 8.34 &4.87   &\textbf{1.23}\\
5 dB  & 16.53 & 11.27 & 9.27 &10.05&6.43  &\textbf{1.52} \\
 0 dB & 25.22 & 17.61 & 13.37 & 14.64&10.16  & \textbf{1.98}\\
  \bottomrule
 \end{tabular}
 }
  \end{threeparttable}
\end{table}

To validate the noise robustness of different methods, the evaluation results measured by \textit{$\ell_1$ distance to model} on the three-component test signal are presented in Table \ref{tab1}, where we change the SNR level of test data from 45  to 0 dB, and 100 trials are implemented at each SNR.  Even though our training dataset only contains signals at SNR = 10 dB, the proposed KL-TFD always achieves a large performance gain compared to other methods especially at low SNRs, which indicates that the kernel learning model is robust to noise.

\vspace{-1mm}
\subsection{TFDs of Real-World Signals}

Although only trained on synthetic data, the proposed KL-TFD method is also validated over real-world data. We now examine the effectiveness of the KL-TFD method compared against WVD, BD, MDD, RGK, and ADTFD on a real-world bat echolocation signal with $K=400$ samples \cite{BOASHASH2018120,BOASHASH201848}. The visualized TFD results of the noiseless bat echolocation signal are shown in the last row of Fig. \ref{fig2}. It is observed that the signal-dependent methods achieve better  performance than signal-independent WVD and BD, i.e., CTs are greatly removed. Moreover, only ADTFD and KL-TFD are capable of gaining the fourth part of the bat signal, whose energy is too low to recognize in most methods. 

\begin{figure}[!t]
\vspace{-3.5mm}
\begin{center}
\hspace{-4.mm}
\includegraphics[width=3.6 in]{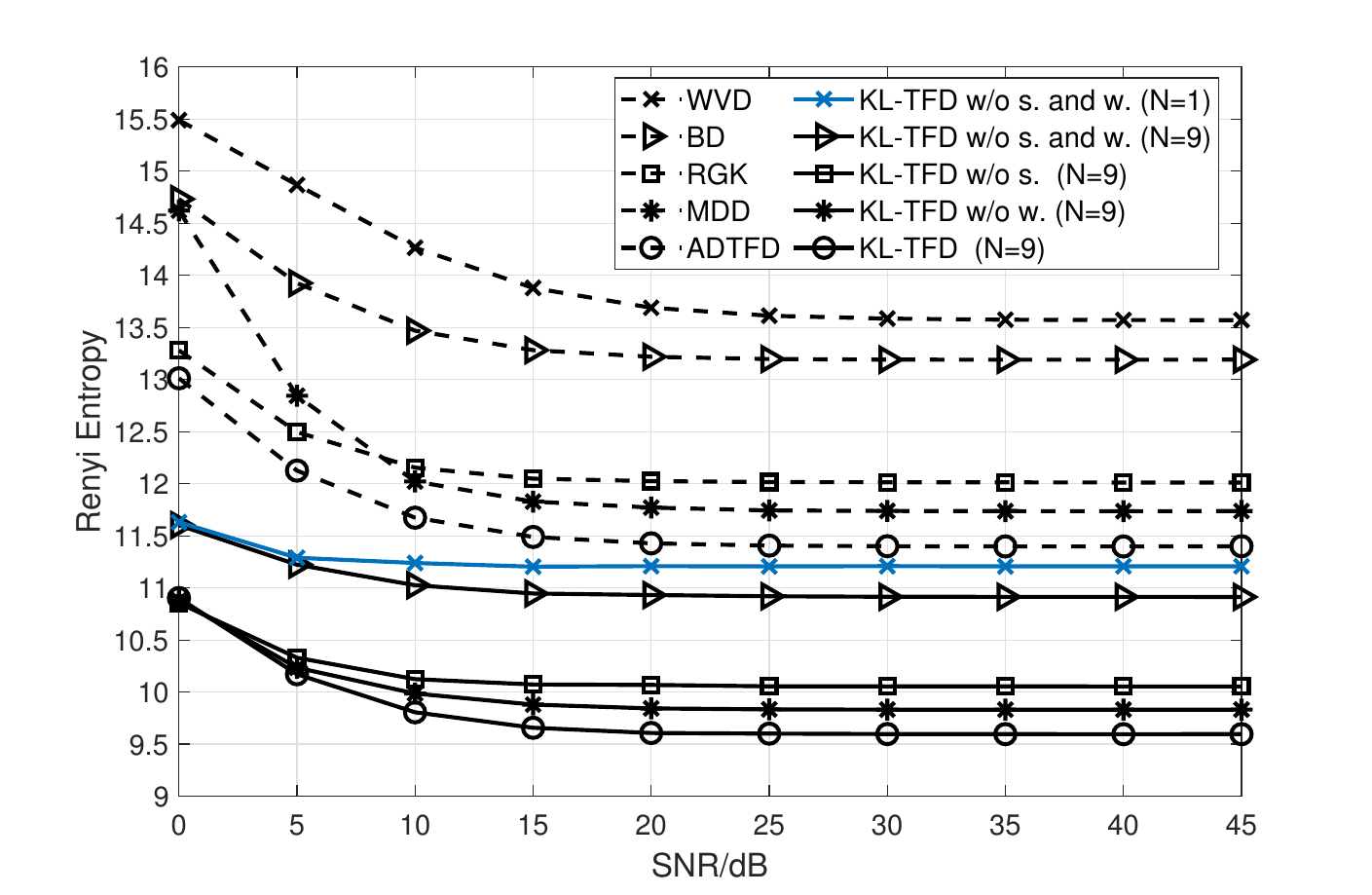}
\vspace{-5mm}
\caption{R\'{e}nyi entropy on real-world bat echolocation signal vs. SNR levels. (s. and w.  denote the skipping operator and the weighted block, respectively)}
\vspace{-4mm}\label{fig3}
\end{center}
\end{figure}

To independently assess the contribution of different parts in  KL-TFD, we conduct a series of  ablation experiments. The quantitative comparison results of different methods measured by \textit{R\'{e}nyi entropy} are presented in Fig. \ref{fig3}.   Three observations can be made. First, the KL-TFD model without (w/o) the skipping operator and the weighted block when $N=1$, i.e., only two 2D Conv blocks left in Fig. \ref{fig1} (A), is trained and the test result is indicated with a blue curve in Fig. \ref{fig3}. Despite the simple network configuration, its performance is still superior  over all the traditional kernel design methods, demonstrating the advantages and feasibility of data-driven kernel learning models. Second, the KL-TFD models  without  the skipping operator and/or the weighted block when $N=9$ are implemented. Note from Fig. \ref{fig3} that both of the skipping operator and the weighted block play important roles in improving performance, and the best result is achieved   when we combine these two parts together. Third, it should be stressed that the trained KL-TFD model on synthetic signals is still effective for actual nonstationary bat signals, and can be generalized to test signals with more number of components.   For other actual signals, e.g., EEG, the training dataset of the KL-TFD can be reconstructed to better learn EEG's characteristics.

\vspace{-0mm}

\vspace{-1mm}
\section{Conclusion}

This letter attempts to generate high-resolution and CT-free TFDs by proposing a data-driven kernel learning model, which is designed as an end-to-end network to replace traditional kernel functions, where two skipping 2D Conv blocks and one weighted block are two dominant parts. Experimental results demonstrate that supervised kernel parameter learning methods using neural network based models for high-resolution TFDs outperform their traditional kernel design counterparts. Ablative  analysis also examines the effectiveness of our proposed KL-TFD model. Future works may consider the signal TF analysis with missing data \cite{amin2019sparsity,Shuimei8735761,KHAN2020107260}.

  \newpage

\bibliographystyle{IEEEtran}
\bibliography{jl_v3}

\end{document}